\documentclass[sigconf]{acmart}
\AtBeginDocument{%
  }

\copyrightyear{2026}
\acmYear{2026}
\setcopyright{cc}
\setcctype{by}
\acmConference[WebSci '26]{18th ACM Web Science Conference}{May 26--29, 2026}{Braunschweig, Germany}
\acmBooktitle{18th ACM Web Science Conference (WebSci '26), May 26--29, 2026, Braunschweig, Germany}
\acmDOI{10.1145/3795766.3799743}
\acmISBN{979-8-4007-2504-3/2026/05}
\usepackage{multirow}
\usepackage{subcaption}

\begin{document}

% \title{The Effect of Calorie Density on Engagement in Reddit's Food Communities}
\title{Reddit's Appetite: Predicting User Engagement with Nutritional Content}


\author{Gabriela Ozegovic}
\affiliation{%
  \institution{Graz University of Technology}
  \city{Graz}
  \country{Austria}}
\email{ozegovic@tugraz.at}

\author{Thorsten Ruprechter}
\affiliation{%
  \institution{Graz University of Technology}
  \city{Graz}
  \country{Austria}}
\email{th.ruprechter@gmail.com}

\author{Denis Helic}
\affiliation{%
  \institution{Graz University of Technology}
  \city{Graz}
  \country{Austria}}
\email{dhelic@tugraz.at}

\renewcommand{\shortauthors}{Ozegovic et al.}

\begin{abstract}
Food communities on online platforms enjoy great popularity among social media users. Due to the far-reaching consequences of food-related content on user eating behavior, recent research has studied the factors that drive user online engagement with food. While most of these studies have focused on visual aspects of food content in social media, only a few initial studies have explored the impact of nutritional content on user engagement. In this paper, we set out to close this gap and analyze food-related posts on Reddit, focusing on the association between the calories and macronutrients of a meal and engagement levels, particularly the number of comments. To that end, we collect and analyze almost half a million food-related posts and uncover differences in nutritional content between engaging and non-engaging posts. Moreover, we train a series of XGBoost models, and evaluate the importance of nutritional content while predicting user engagement and how posts will resonate with the community. We find that nutritional features improve the baseline model’s accuracy by almost 5\%, with a positive contribution of calorie density towards the prediction of engagement, suggesting that higher nutritional content is associated with higher levels of user engagement in food-related posts. Our results provide valuable insights for the design of more engaging online initiatives aimed at, for example, encouraging healthy eating habits.
\end{abstract}




%\received{31 January 2025}

\maketitle

\newpage
\section{Introduction}
Nowadays, users increasingly share food-related content online by posting recipes, meal plans, or dietary advice.
While over one million recipes are already available on the Web \cite{salvador_learning_2017}, social media platforms further amplify this trend. 
For example, as of November $2025$, Instagram alone has more than $548$ million posts with the hashtag ``food.''
Recently, several studies have analyzed factors driving this substantial user engagement with food-related online postings. While individual post features such as positive language and emotions are, in general, associated to increased level of user activity\cite{barklamb_learning_2020}, the engagement factors related to food content are typically more intricate.
For example, temporal aspects along with the reputation of the author of the posting, strongly affect how people engage with food content \cite{rokicki_how_2017}. 
In addition, visual appeal of the food is also associated with the user engagement levels \cite{lee_visual_2023}. 

High levels of user involvement with food on social media raise the question of the health implications of this activity. For instance, it is still mostly unclear whether users tend to engage more with food posts that promote healthy eating practices or with posts that, for instance, contain high-calorie meals \cite{garaus2021unhealthy}.
Recently, some initial studies have explored the association between nutritional content and online engagement, showing a positive correlation between engagement and nutritional profile of food \cite{pancer_content_2022}. 
In that study, the authors analyzed $700$ Facebook posts featuring Buzzfeed's Tasty videos, showing that posts featuring calorie-dense meals receive more likes, shares, and comments.
These small-scale studies analyzing a few hundred food posts give fruitful insight into the health implications of high user engagement with certain food-related content online.
However, it is unclear whether these findings generalize to a global online community that generates large amounts of food content and attracts vast amounts of user attention.

In this paper, we build upon those previous studies by examining engagement with food-related posts on Reddit, in particular on r/Food, an online food sharing community. To that end, we compute nutritional content (calories and macronutrients per $100$g) from food post titles and investigate the association of these attributes with user engagement. 
With our paper, we expand the previous work in two important ways. 
First, we apply a robust embedding-based method for estimating nutritional content from the food post title only. 
Second, we conduct a large-scale analysis of over half a million Reddit food-related posts and analyze the factors of user engagement in these posts.
Particularly, adopting user comments as a measure of engagement, 
we focus our analysis on the top 1\% of posts by the number of comments. Using those posts, we quantify the association between nutritional factors and engagement by training a series of XGBoost classifiers \cite{chen_xgboost_2016} for predicting high engagement posts. To isolate the relation between nutritional content and engagement we control for several non-food-related features, such as seasonality or user tenure, by including them as predictors in our classifiers. We use SHAP values \cite{lundberg_unified_2017} for a detailed explanation of the predictive power of nutritional content. 

We find that, even after controlling for post, user, or temporal features, posts featuring more nutrient-dense meals are positively associated with engagement. In particular, these posts are more likely to obtain comments and resonate with the community. 
In total, nutritional features improve the prediction performance of the baseline XGBoost models by almost 5\%, indicating a robust association between nutritional content and engagement in food-related posts. However, we find similar performance improvements (around 5\%) over the controls when using significant discriminative words appearing in the title such as ``cheese'', ``pizza'', or ``chocolate'', and even stronger performance gains (almost 17\%) when using visual features of the images included in the postings corroborating findings from previous studies \cite{lee_visual_2023}. Hence, our results suggest an intricate association between nutritional content, usage of popular meals in the titles, food visual appearance, and user engagement in food-related Reddit posts.

Our work provides a deeper understanding of factors associated with user engagement, particularly the nutritional content of food.
Given the high obesity rates\footnote{\url{https://www.who.int/news-room/fact-sheets/detail/obesity-and-overweight}}, knowledge of how users interact with food content online is crucial.
The explainability of our models via SHAP values reveals the structure of posts that receive high engagement rates.
As a result, this enables the design of more engaging online initiatives aimed at encouraging healthy eating habits.
In addition, our approach to calculating the nutritional content of a meal from just the textual description can be used in dietary education, helping people understand the nutritional profile of their meals.
Alongside promoting greater nutritional awareness, our research can potentially contribute to the broader societal effort to combat obesity and foster healthier online food discussions.
In addition, we publish all of our code and data\footnote{\url{https://github.com/gabrielaozegovic/reddits-appetite}}.

\section{Related Work}
\noindent\textbf{Food preferences and food choices.} Physical food features affect how individuals respond to food. 
Brain fMRI studies show that calorie-dense food provokes palatable and satiating feelings, whereas low-calorie food provokes hunger \cite{killgore_affect_2006}.
Moreover, social influence plays a key role in food choices; individuals often choose healthier options when their eating partners do as well \cite{gligoric_formation_2021}. 
This reflects the broader impact of social cues on food decisions. 
For example, people mimic a thin person's large portion size but opt for smaller portions when the person appears obese \cite{mcferran_ill_2010}.
Similarly, students are more likely to purchase a food item if the person ahead of them buys it \cite{gligoric_food_2024}. 
Social networks can amplify these effects; a person is up to $57$\% more likely to become obese if someone in their close social circle does \cite{christakis_spread_2007}. 
Such traditional studies, usually requiring active participation, can suffer from small sample sizes (e.g., n=$59$, n$=139$) \cite{houben_guilty_2010, serrano-gonzalez_developmental_2021} and may miss real-world nuances.
Therefore, researchers already studied online user behavior related to food.
For example, individuals beginning a diet tend to search for lower-calorie meals
online \cite{west_cookies_2013}. 
Likewise, tweets about high-calorie foods correlate with state-wide obesity rates in the US \cite{abbar_you_2015}. 

Similar to other online studies, and in contrast to controlled ones, we use online data to explore food preferences on a large scale, aiming to gain insights from a larger and more diverse sample.

\noindent\textbf{Food and social media.}
Social media engagement is complex, influenced by various factors such as content or algorithms \cite{yan_evolution_2024}. 
Visual and persuasive content, including emotions and humor, typically generates higher engagement \cite{lee_effect_2013}. 
Specifically on Reddit, images and captions are highly predictive of user engagement \cite{hessel_cats_2017}.
Related to food posts, Philp et al. \cite{philp_predicting_2022} found that users are more likely to interact with posts featuring food with a more typical appearance.
Also, visually appealing food increases purchase intentions and promotes healthier food choices \cite{lee_visual_2023}. 
Further, while Instagram engagement is driven by longer captions and food health information \cite{barklamb_learning_2020}, 
Turnwald et al.  showed that Instagram food posts from influencers with less healthy food ratings received more engagement, and this appeal extends beyond sponsors and advertisements \cite{turnwald_nutritional_2022}.

Nevertheless, the association of nutritional content with engagement remains under-investigated, despite its potential implications for promoting healthier eating behaviors.
Therefore, our study focuses on calorie and macronutrient content to explore how nutritional information is associated with engagement. 
We conduct an observational study of user engagement in settings where users interact directly with each other, rather than with brands or influencers. 
These user-to-user interactions offer more authentic insights into behavior, free from the influence of ads and marketing.

\noindent\textbf{Estimating nutritional content.}  
Discussions about calorie content on Twitter increased following the U.S. federal calorie labeling law \cite{hswen_federal_2021}, reflecting public interest in nutritional information.
Yet, a study of over $1,000$ Instagram food posts revealed that over $90$\% of the posts lack nutritional information \cite{kabata_can_2022} and that fewer than $4$\% of images under diet hashtags contain nutrition data \cite{lister_what_2024}.
Recently, some initial studies estimated the nutritional information from social media postings. For example, Turnwald et al. \cite{turnwald_nutritional_2022} manually labeled food using images and captions and matched them to entries in the food database.
In addition, researchers have calculated calorie information by performing keyword matches between posts and entries in food databases or nutritional information websites \cite{sharma_measuring_2015, abbar_you_2015}. 
While convenient, this method is susceptible to issues arising from inconsistent phrasing and may require manual verification.

For our study, we compute calories using text embedding techniques, allowing us to aggregate multiple similar meals. 
This makes our calorie and macronutrient estimation approach more robust to variations in user-generated post titles.

\section{Materials and Methods}

\begin{figure*}[t]
    \centering
    \scalebox{1}{ % adjust to e.g. 0.9 if you need more space
    \begin{minipage}{\textwidth}
    \begin{subfigure}[t]{0.348\textwidth}
        \centering
        \vspace{0pt}
          \includegraphics[width=\textwidth, height=3.85cm]{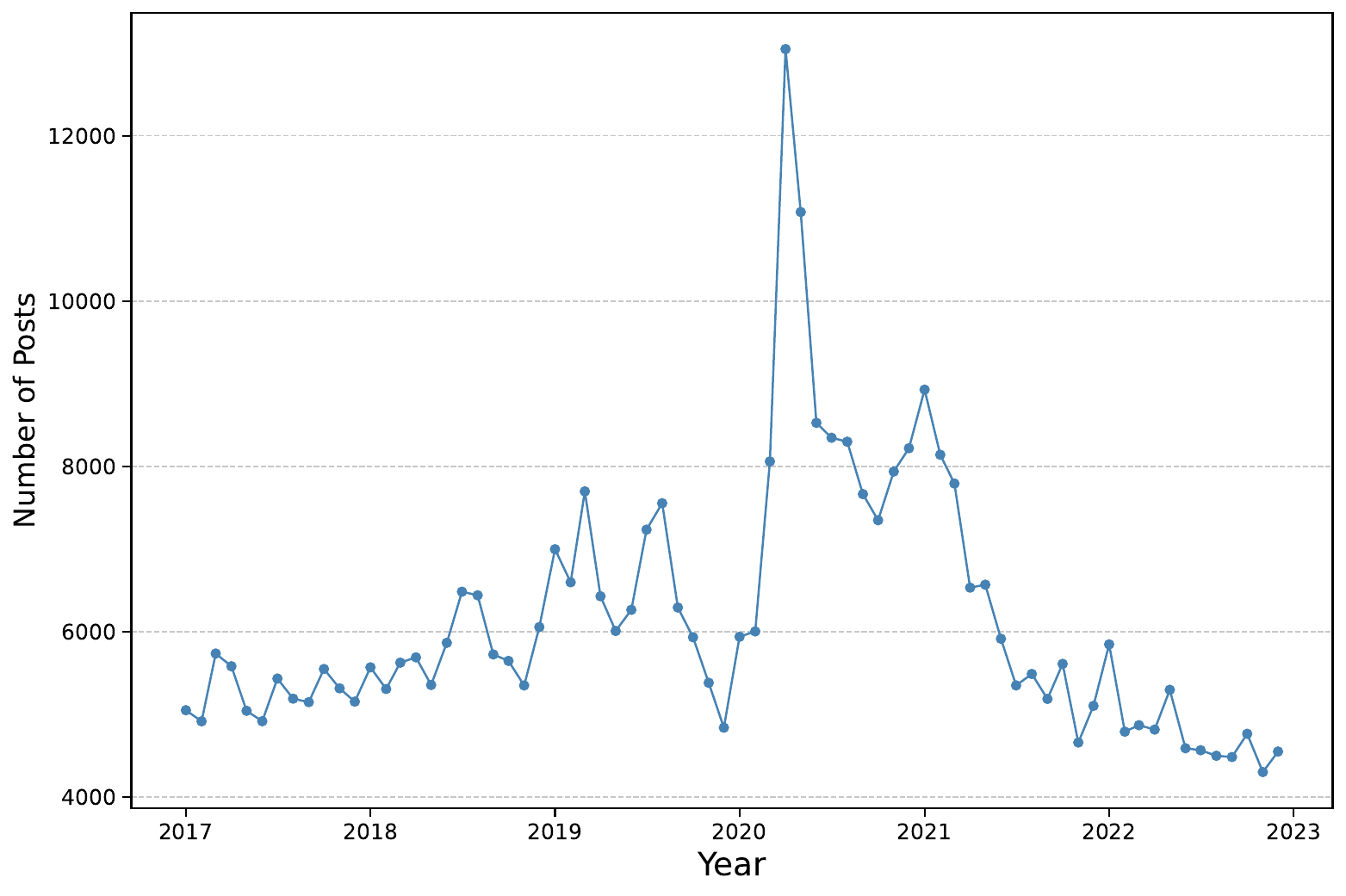}
          \caption{Number of Posts by Year}
          \Description{Chart shows number of posts by year, with posts increasing before COVID-19, peaking during the pandemic, and returning to pre-pandemic levels after.}
        \label{fig:posts_per_year}
    \end{subfigure}
    \hfill
    \begin{subfigure}[t]{0.372\textwidth}
        \centering
        \vspace{0pt}
          \includegraphics[width=\textwidth, height=4cm]{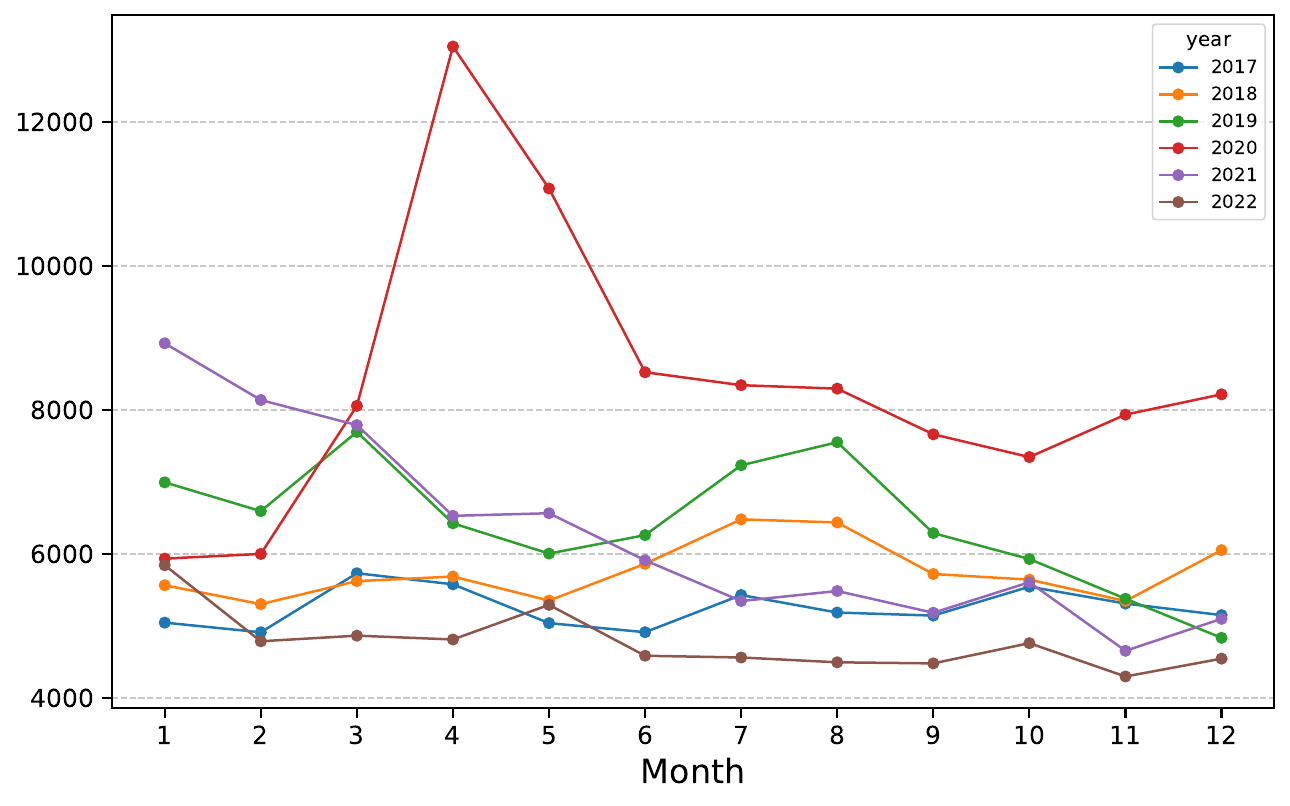}
          \caption{Number of posts by Month}
          \Description{Chart shows number of posts by month, usually between $5,000$ and $8,000$ posts per month, peaking between March and June $2020$ during the pandemic.}
        \label{fig:posts_per_month}
        \end{subfigure}
    \hfill
    \begin{subfigure}[t]{0.24\textwidth}
        \centering
        \vspace{0pt}
        \includegraphics[width = \textwidth, height=4cm]{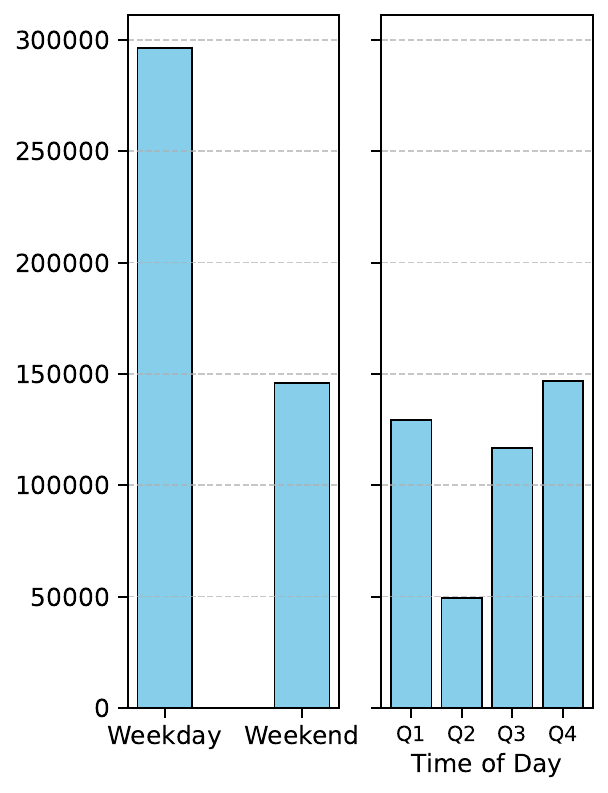}
        \caption{Number of Posts by Day Type and Time of Day Quartiles}
        \Description{Chart shows the number of posts by day type (Weekday or Weekend) and time of day quartiles.}
        \label{fig:posts_weekend}
        \end{subfigure}

    \medskip

    \centering
    \begin{subfigure}[t]{0.348\textwidth}
        \centering
        \vspace{0pt}
          \includegraphics[width=\textwidth, height=3.85cm]{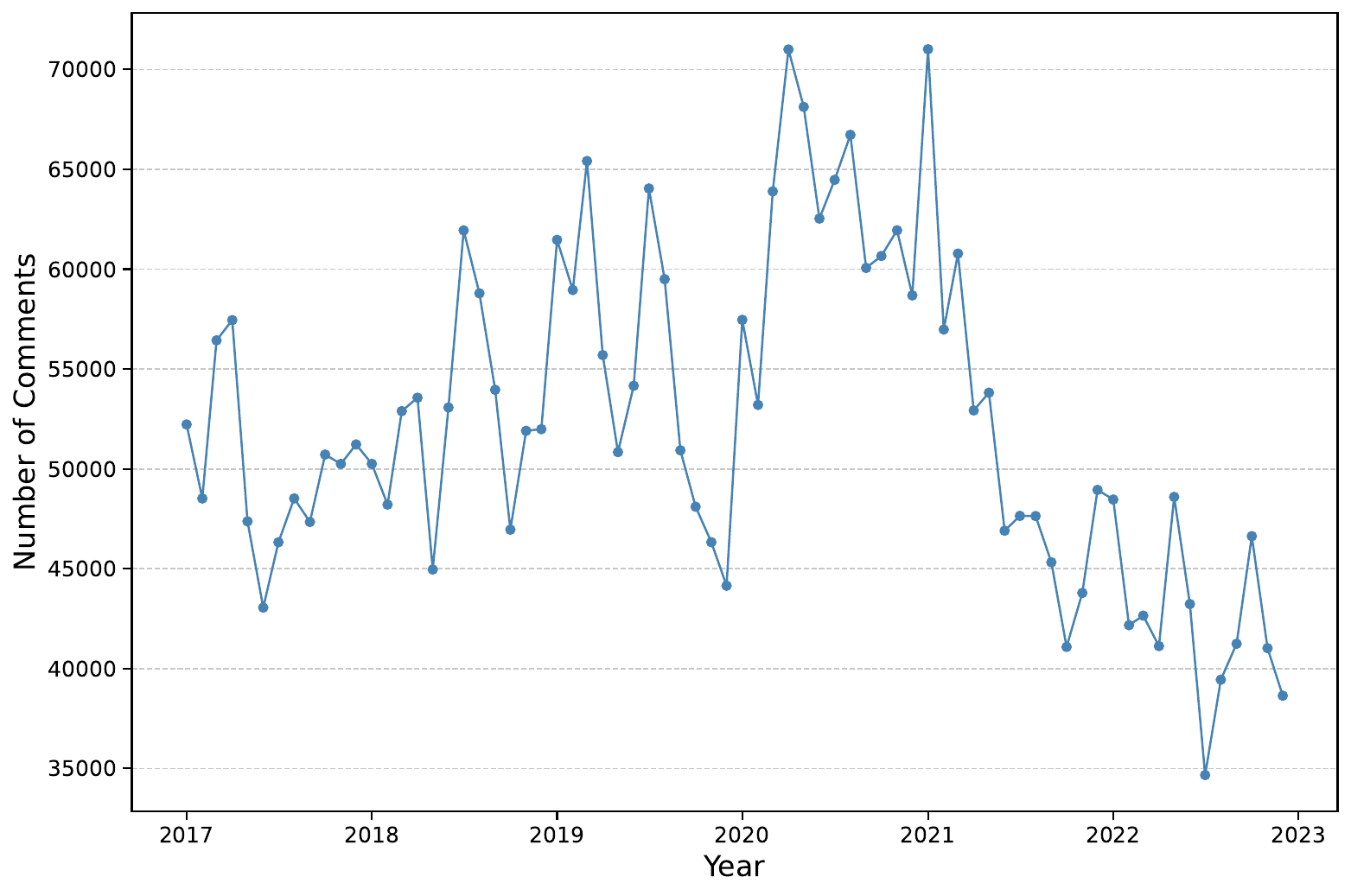}
          \caption{Engagement Level by Year}
          \Description{Chart shows engagement level by year, peaking during the pandemic.}
        \label{fig:comm_per_year}
    \end{subfigure}
    \hfill
    \begin{subfigure}[t]{0.372\textwidth}
        \centering
        \vspace{0pt}
          \includegraphics[width=\textwidth, height=4cm]{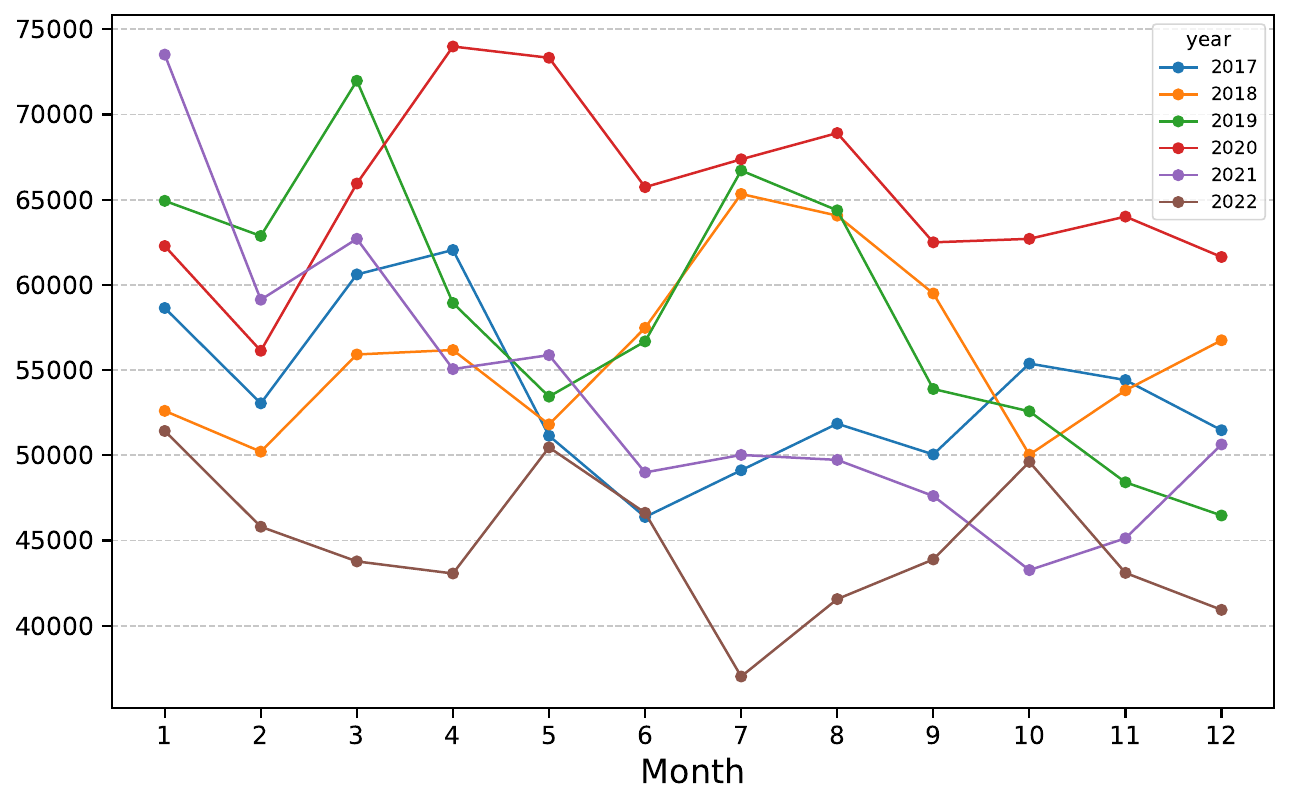}
          \caption{Engagement Level by Month}
          \Description{Chart shows engagement level by month.}
        \label{fig:comm_per_month}
        \end{subfigure}
    \hfill
    \begin{subfigure}[t]{0.24\textwidth}
        \centering
        \vspace{0pt}
        \includegraphics[width = \textwidth, height=4cm]{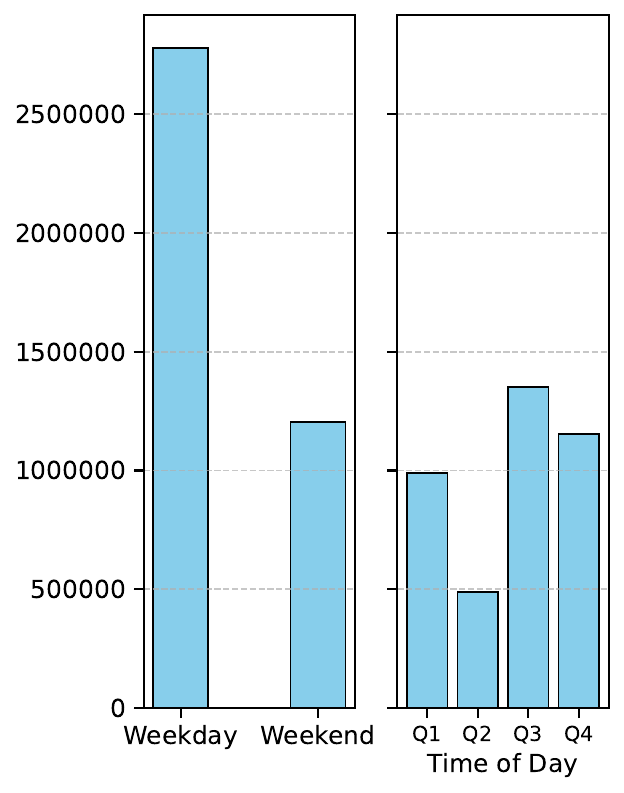}
        \caption{Engagement Level by Day Type and Time of Day Quartiles}
        \Description{Chart shows engagement levels by day type (Weekday or Weekend) and time of day quartiles.}
        \label{fig:comm_weekend}
    \end{subfigure}
    \end{minipage}
    }
    \caption{
    \textit{Posts and comments in r/Food over time.} We present how postings and comments developed from $2017$ until $2023$ across different temporal scales, including yearly, monthly, weekly, and daily trends.
    In (a) we present the number of posts over the years. We observe a positive trend before the COVID-19 pandemic, with a noticeable peak during the pandemic, and a drop afterwards to pre-pandemic levels.
    Monthly posting activity in (b) is rather consistent except for a peak between March and June $2020$ during the pandemic.
    In (c) we observe that more posts are created on weekdays than on weekends (left) and that most posts are created in the afternoon in the eastern USA (Q4, right). The bottom row shows the same diagrams for comments.
    In (d) we observe a gradual increase in commenting activity over time, with the highest activity levels during the pandemic and a sharp drop after the pandemic.
    This observation is also reflected in (e), where we see constant high levels of comments in $2020$. We also see a seasonal spike in January, possibly due to the holiday season.
    In (f), comments mirror posting activity, with more comments over the weekdays (left). On the other hand, the peak in comments is in the morning (Q3, right).
    }
    \Description{Six charts arranged in two rows illustrate posts and comments in the r/Food subreddit from $2017$ to $2023$. Top Row (Posts): Chart (a) shows yearly posts increasing before COVID-19, peaking during the pandemic, and returning to pre-pandemic levels after. Chart (b) shows monthly consistency except for a major spike from March to June 2020, during the pandemic. Chart (c) indicates higher posting volume on weekdays and a daily peak in the afternoon (Eastern USA time). Bottom Row (Comments): Chart (d) shows a gradual rise in comments that peaked during the pandemic before a sharp decline. Chart (e) displays constant high levels in 2020 and recurring spikes every January. Chart (f) shows more comments on weekdays with a daily peak in the morning.}
    \label{fig:temporal}
\end{figure*}

\subsection{Dataset}

\noindent\textbf{Reddit.}
Reddit is an online platform consisting of multiple discussion communities, called ``subreddits''. 
Typically, subreddits are focused on a specific topic, and users write posts or comment on existing posts, forming a shared interest-centric community. 
Each subreddit has its own rules and guidelines on what to include in the post title and body, formatting instructions, or general instructions on communication tone and how to particiapte in that community. 

\noindent\textbf{Food subreddit.}
In this paper, we focus on r/Food, a subreddit dedicated to sharing meals. As of November $2025$, it is the $22$nd largest subreddit, with around $24$ million subscribers\footnote{\url{https://www.reddit.com/best/communities/1/\#t5_2qh55}}. 
In particular, users post meals, following the rules of the subreddit: the post title must describe the meal.
For example, typical posts have titles such as ``[Homemade] Roasted Pork Belly'' or ``[I Ate] Salted Caramel Pork Floss Ice Cream''.
Additionally, each post must include an original image of the meal, taken by the user who creates the post.
These rules ensure consistency across user posts and their focus on food. Even though the rules slightly changed over the years, the meal name had to be always included in the post title.

\noindent\textbf{Data collection.}
We collect data with Pushshift, a service that conducts large-scale crawls of Reddit \cite{baumgartner_pushshift_2020}.
We retrieve all submissions ($594,842$ posts) from r/Food subreddit from January $2017$ up to the end of December $2022$. 
For each post, we collect the number of comments the post received as a basic measurement of community engagement.
In addition, we collect further post information such as username or submission time, and if present the post image.

\noindent\textbf{Ethical Considerations.}
We use publicly available submissions and comments for our analysis. Our data collection and usage comply with the Reddit's terms of service\footnote{\url{https://www.reddit.com/r/reddit.com/wiki/api-terms/\#wiki_3.__fees.3B_restrictions_on_use}}, and do not impose ethical risk to users involved in this subreddit. We avoid direct user interactions and ensure anonymity by focusing on aggregate data rather than individual users.

\noindent\textbf{Preprocessing.}
First, we remove empty and deleted posts, as the community does not engage with such posts. Next, we remove duplicate posts, which we define as those made by the same user with the same title within five minutes. 
In the remaining $513,044$ posts, we clean up titles by removing special characters and emojis.
Lastly, we only keep posts with a valid image, resulting in $447,863$ posts.

\begin{figure*}[t]
    \centering
    \begin{subfigure}{0.25\textwidth}
        \centering
        \includegraphics[width=\textwidth]{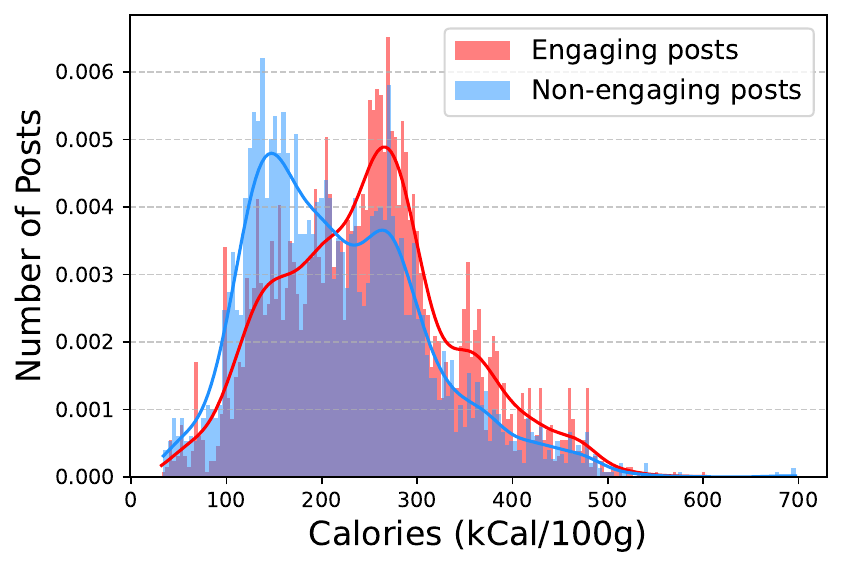}
        \caption{Calorie Densities}
        \Description{Density plot shows calorie distributions.}
        \label{fig:rq2_calorie}
    \end{subfigure}
    \hfill
    \begin{subfigure}{0.24\textwidth}
        \centering
        \includegraphics[width =\textwidth]{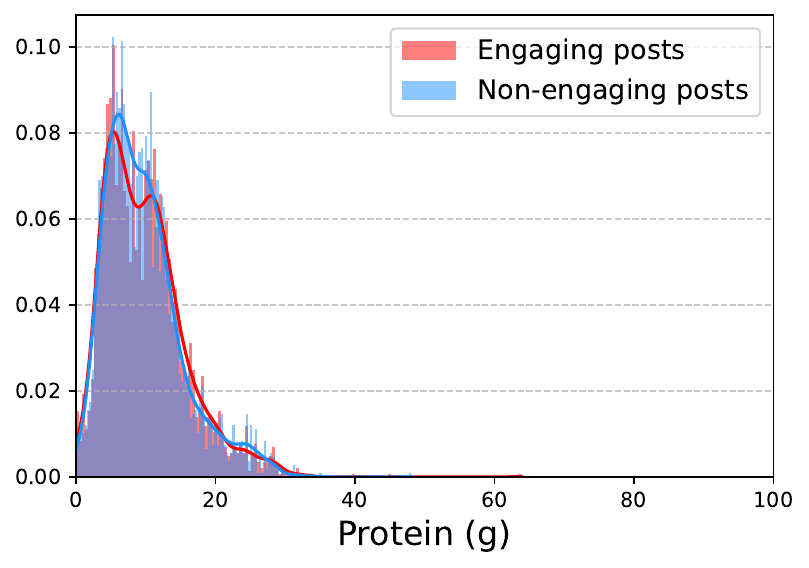}
        \caption{Protein Densities}
        \Description{Density plot shows protein distributions.}
        \label{fig:rq2_protein}
        \end{subfigure}
    \hfill
    \begin{subfigure}{0.24\textwidth}
        \centering
        \includegraphics[width =\textwidth]{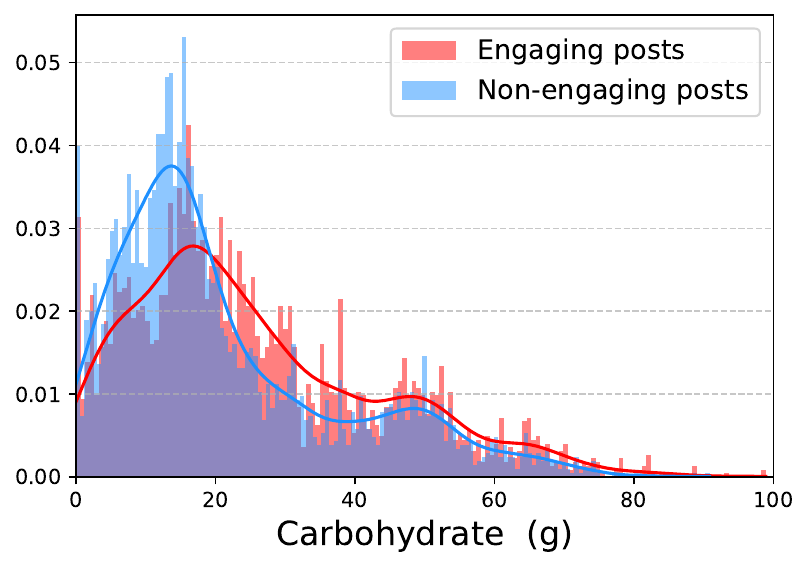}
        \caption{Carbohydrate Densities}
        \Description{Density plot shows carbohydrate distributions.}
        \label{fig:rq2_carb}
        \end{subfigure}
    \hfill
    \begin{subfigure}{0.24\textwidth}
        \centering
        \includegraphics[width = \textwidth]{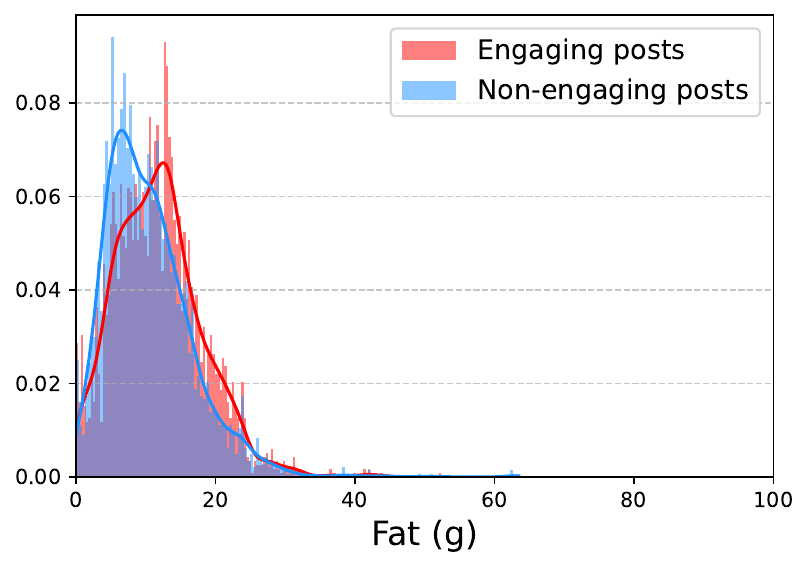}
        \caption{Fat Densities}
        \Description{Density plot shows fat distributions.}
        \label{fig:rq2_fat}
        \end{subfigure}

    \caption{
    \textit{Nutritional content distribution of food in r/Food posts}. We illustrate the distribution of calories (a) and macro-nutrients (b--d) per $100$g of food, across meals in engaging (red) and non-engaging (blue) posts.
    The calorie content is measured in kCal per $100$g, while macro-nutrients are measured in grams as fractions of $100$g total.
    We observe that the majority of posts fall within the moderate calorie range, between $100$ and $300$ kCal.
    Distribution disparities are prominent when comparing engaging vs. non-engaging posts. In particular, non-engaging posts peak at around $150$ kCal, while posts attracting user engagement peak at around $300$ kCal (a).
    We observe a sharp cut-off in the protein (b) density, with most posts having less than $20$g of protein, suggesting a prevalence of low-to-moderate protein meals.
    Carbohydrates (c) span over a wider range. While most posts have less than $30$g, there is a consistent amount of carb-rich food as well, as indicated by the long tail in their distributions.
    Fat (d) distribution peaks around $10$-$15$g, with most posts containing moderate fat content.
    However, distributions of all macronutrient densities are shifted to the right for engaging posts as compared to non-engaging posts.
    }
    \Description{Four density plots compare the nutritional content per 100g of food in engaging and non-engaging posts. Across all four charts, the distributions for engaging posts are shifted to the right, indicating higher caloric and macronutrient densities.}
    \label{fig:macros}
\end{figure*}

\subsection{Nutritional Content Estimation}
To calculate the nutritional content of each meal, we use USDA FoodData Central database \cite{mckillop_fooddata_2021} including Foundation Foods, SR Legacy, and The Food and Nutrient Database for Dietary Studies. 
We compute the nutritional content from the titles of Reddit posts by adapting the NutriTransform method \cite{ruprechter_2025}.
In particular, we compute sentence embeddings (\textit{sentence-transformers/all-mpnet-base-v2}) \cite{reimers_sentence-bert_2019} for both Reddit post titles and the food database items. 
Using these embeddings, we compute the cosine similarity between a given Reddit post and all meals from the food database. We then select the five closest matches to the Reddit post, given that they exceed a similarity threshold of $0.6$. To determine that threshold, we sample $5,000$ Reddit posts and compute their similarities to all $11,801$ food database items, and set the threshold to the median of the distribution of $99.9$th quantile similarities. We test the robustness of this similarity threshold by making additional computations with varying quantiles (e.g., $99.99$, $99$, $95$) and find no significant impact of the alternative similarity thresholds on our results. After selecting the five most similar meals from the database, we extract their calorie and macronutrient information and aggregate these values by computing a similarity-weighted mean. This provides an estimate of the nutritional content for a given post.
As the entries in the USDA FoodData database are given per $100$g of a meal, all calculated calorie and macronutrient information also represent densities per $100$g of food.

Using our method, we compute the nutritional information for $326,388$ different meals, as multiple posts can contain the same meal (e.g., $1,591$ posts have the title ``Pizza'').
We exclude posts for which we did not find any suitable matches in the food database, i.e., that are below the similarity threshold.
As a post-processing step, we check for potential outliers, which we define as meals with less than $32$ calories (equivalent to $100$g of strawberries) or more than $717$ calories (equivalent to $100$g of butter). Moreover, we remove all meals where protein, carbohydrate or fat estimates are over $100$g.
After this final filtering step, we have a total of $325,086$ meals in $442,371$ posts that we use for further analysis.

\noindent\textbf{Validation.} We validate our method on two different datasets: (i) the labeled recipe dataset by West et al. \cite{west_cookies_2013}, and (ii) Google Nutrition5k dataset \cite{thames_nutrition5k_2021}. The first data set contains $9,859$ recipe titles with corresponding nutritional information, making it similar to our target data set. Following the same steps as in our main approach, we estimate nutritional content for $9,144$ recipes. Comparing our estimates to the ground truth from the dataset we obtain mean absolute error (MAE) of $60.8$ and root mean squared error (RMSE) of $85.6$ kCal. The second dataset Nutrition5k contains images of $4,768$ plates of food, along with the total meal mass, nutritional information, and a list of ingredients with their weights. Since food names are not provided, we construct them using the ingredients. Specifically, we create a ``food title'' by concatenating the top five ingredients by their weight, given that each ingredient accounts for at least $5\%$ of the total meal weight. After removal of title duplicates, we end up with $2,342$ unique meals. Using those meals, we are able to estimate the nutritional information for $2,299$ meals, resulting in MAE of $83.5$ and RMSE of $122.9$. We attribute lower performance on this dataset to the manual construction of food titles.

\subsection{Explorative analysis}
\noindent\textbf{Users.}
A total of $152,362$ unique users contributed posts to the subreddit, with $61.7$\% posting only once.
The most active user made $952$ posts. Typically, more active users have more experience, and the community engages strongly with their posts \cite{rokicki_how_2017}. 
In our dataset, the top $5$\% of users ($6,888$ users) have at least $10$ posts each.

\noindent\textbf{Comments.}
The mean number of comments per post is nine, with a standard deviation of $43.4$, indicating significant variability in comment counts. In total, $338,747$ ($76.6$\%) posts received at least one comment.  
The maximum number of comments on a post is $2,447$, while the median is only two, and the third quartile is just six comments, indicating a strongly skewed distribution. Although the community engages with the majority of the posts, highly engaging posts (top 1\%) receive all a minimum of $215$ comments.

\noindent\textbf{Scores.}
Each Reddit post has a score, defined as the difference between the community's ``upvotes'' and ``downvotes''. 
The mean score is $243$, with a standard deviation of $1,739.58$, indicating high variability. The median score is only $22$, signaling again a skewed distribution where most posts receive modest scores while the highest score is $70,308$.
In this paper, we do not use score as an engagement metric and opt for comments, which require more user effort. In addition, score and comment count are strongly positively correlated ($\rho = 0.87, p < 0.001$), indicating that comments are a comprehensive representation of engagement. %Spearman

\noindent\textbf{Temporal characteristics.}
In Figure \ref{fig:temporal}, we depict the temporal development of user activity and user engagement in r/Food. The number of posts steadily increased over time (Fig. \ref{fig:posts_per_year}), peaking in $2020$, likely due to the COVID-19 pandemic and its associated lifestyle changes, e.g., increased interest in food and consumption of more meals at home \cite{gligoric_population-scale_2022}. After a brief increase in early $2021$, posts declined, falling below pre-pandemic levels.

Monthly post counts across years (Fig. \ref{fig:posts_per_month}) show a similar yearly pattern, with a spike in March $2020$ corresponding to the pandemic onset. Figure \ref{fig:posts_weekend} shows more posts are made on weekdays ($296,349$) than weekends ($146,022$), peaking in the afternoon (Q4, eastern USA, likely reflecting users' lunchtime) with $33.2$\% of posts, followed by evening (Q1, $29.2$\%), morning (Q3, $26.4$\%), and night (Q2, $11.2$\%). Time is interpreted using EST, as most Reddit traffic comes from the US (cf. Reddit traffic as of March 2024\footnote{\url{https://www.statista.com/statistics/325144/reddit-global-active-user-distribution/}}).

The bottom row of Figure \ref{fig:temporal} shows the same analysis for comments, categorized by the post date (e.g., all comments on a June $2020$ post are treated as June $2020$). Comment activity gradually increases until $2020$, followed by a sharp drop and peak aligned with the pandemic. After early $2021$, comments declined below pre-pandemic levels. No clear monthly seasonality is observed (Fig. \ref{fig:comm_per_month}). Comments are more frequent on weekday posts (Fig. \ref{fig:comm_weekend}), with morning posts receiving the most comments (Q3, $33.9$\%), followed by afternoon (Q4), evening (Q1), and night (Q2).

\noindent\textbf{Tags.}
According to the current subreddit rules, each post must include a tag indicating the context of the meal: whether the user prepared it at home, whether the user works in the food industry and prepared it, or whether the user purchased it without personal preparation. The majority of meals, $74.5$\%, were prepared at home by the users, while $19.7$\% were purchased without any preparation, and $1.5$\% were prepared by food industry professionals. 
The remaining $4.3$\% of posts lack an eligible tag, most likely due to earlier subreddit policies of not enforcing the tag structure.

\noindent\textbf{Engagement levels.}
To obtain a strong contrast between engagement levels, we portray engagement by the number of comments and define posts in the top $1$\% by comment count as engaging ($3,003$ posts), and consider those with zero or one comment as non-engaging ($157,470$ posts).
As we want to control for visual features, we consider only posts with images available.
Slight variations in the definition of low engagement, such as considering only posts without comments, posts with just one comment, or posts with up to five comments, do not impact the results. To obtain balanced classes for our prediction experiment (cf. Sect. \ref{sec:prediction}), we randomly sample $3,003$ non-engaging posts, resulting in $6,006$ posts for further analysis. 

\noindent\textbf{Nutritional content analysis.}
We show the distributions of macronutrient content of posts in Figure \ref{fig:macros}. 
Specifically, we compare the nutritional content distributions of engaging and non-engaging posts. 
While the majority of posts fall within the moderate calorie range, from $100$ to $300$ kcal per $100$g of food, we observe a clear difference between posts with and without engagement.
First, a significant difference in means is observed in the calorie distribution ($p < 10^{-53}$).
Most non-engaging posts contain meals with fewer than $150$ kCal, while the majority of engaging posts contain meals with around $250-300$ kCal (Fig. \ref{fig:rq2_calorie}).
Above $220$ kCal, the number of posts that receive user engagement is constantly higher than the number of non-engaging posts.
Further, posts without engagement show a peak at around $5$g of protein, with a gradual decline in posts as protein content increases.
In contrast, posts with engagement exhibit a spike at around $5$g of protein, followed by another increase at just over $10$g of protein. 
While both types of posts tend to feature low-protein meals (up to $20$g), higher-protein meals are more often found in engaging posts (Fig. \ref{fig:rq2_protein}). However, there is no significant difference in means between the protein distributions ($p = 0.1$).
Similarly, both low-carbohydrate and low-fat values are associated with non-engaging posts (Fig. \ref{fig:rq2_carb} and Fig. \ref{fig:rq2_fat}).
Conversely, higher carbohydrate and fat values are linked to posts that get higher engagement rates within the community.
After around $20$g of carbohydrates, the number of engaging posts continually exceeds the numberf of non-engaging posts. Likewise, posts with more than $15$g of fat regularly receive more engagement, with the two groups becoming roughly equal at around $25$g.
The means of both these macronutrients are significantly different between engaging and non-engaging posts ($p < 10^{-27}$).

\section{Predicting Engagement}
To determine whether macronutrients are predictive of user engagement with Reddit food posts, we conduct a binary classification experiment using an XGBoost classifier. We start by extracting five feature sets, including nutritional densities, food descriptors, textual and visual features, and a set of control features.

\subsection{Features}
 
\noindent\textbf{Nutritional content.}
Our primary focus is the nutritional content of meals, including calories (kCal per $100$ grams), protein, carbohydrates, and fat, measured in grams per $100$ grams of a meal. 

\noindent\textbf{Food descriptors \& categories.} 
We describe food according to taste, texture, and preparation method and by a food category (e.g., main dish, dessert, fast food, healthy, etc.) using sets of keywords.
For each descriptor, we first manually select three common keywords and extend these lists by two additional keywords from ChatGPT, which we use for its ability to suggest diverse yet commonly used terms.
For food categories, we manually identify several common keywords and expand the dessert list using suggestions from ChatGPT.
We show the complete list of food descriptors, categories, and their corresponding keywords in Table \ref{table:food_types}.

\begin{table}[b]
\centering
\caption{\textit{Food descriptors \& categories.} Keywords used to classify meals by identifying specific terms in post titles.}
\Description{Table showing keywords used as food descriptors and food categories.}
\resizebox{\columnwidth}{!}{%
\begin{tabular}{ll}
\toprule
Food descriptors    & Definition                                                       \\ \hline
Preparation method  & grilled, fried, baked, boiled, steamed                           \\
Taste descriptors   & savory, sweet, spicy, rich, salty                                \\
Texture descriptors & creamy, crispy, tender, juicy, crunchy                           \\ \hline \hline
Food categories     & Definition                                                       \\ \hline
Main dish           & pasta, casserole, roast, chicken, stirfry                        \\
Dessert             & \vtop{\hbox{\strut cake, custard, pudding, cookie, pancake, }\hbox{\strut waffle, muffin, biscuit}}  \\
Fast food           & pizza, burger, burrito                                           \\
Healthy             & soup, salad                                                      \\
Plant-based          & vegan, vegetarian, veggie                                        \\ 
Pastry              & bread, croissant                                                 \\
\bottomrule
\end{tabular}
}
\label{table:food_types}
\end{table}

Using the descriptor and category keywords, we check whether post titles contain those keywords by performing string matching.
If we find a match, we mark the corresponding descriptor or category as present.
For food descriptors, we use individual keywords as features. For example, for preparation methods, we check the presence of each keyword (e.g., ``grilled'').
On the other hand, for food categories, we use the main categories as features, and the keywords only to identify matches. In particular, if ``soup'' or ``salad'' is present in the title, we mark the post as ``Healthy''. 
This approach also works with posts that belong to multiple categories (e.g., a chicken salad can be categorized as both a main dish and healthy).

\begin{figure}[t]
        \centering
        \scalebox{1}{
        \includegraphics[width=\linewidth]{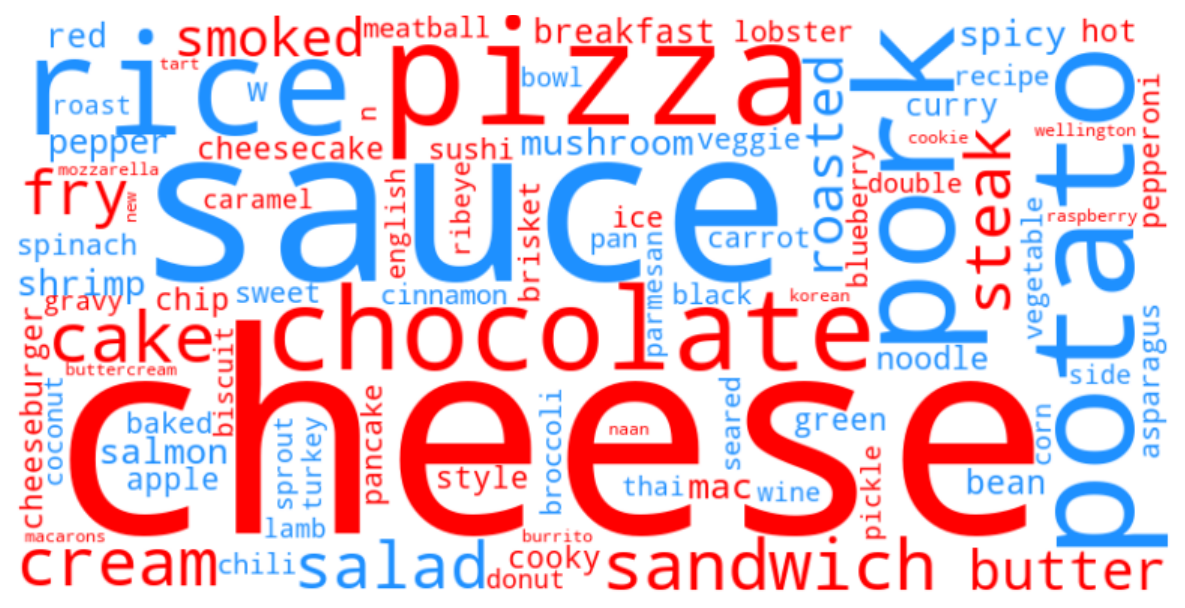}
        }
    \caption{\textit{Engagement discriminators from post titles.} We present discriminative words used significantly differently in engaging and non-engaging posts as word clouds. The red color indicates words more frequently used in posts with engagement. Blue color represents discriminative words more frequently used in posts without engagement. The size of each word reflects its frequency within the respective group.}
    \label{fig:worclouds_both}
    \Description{Word cloud showing engagement discriminators from post titles.}
\end{figure}

\noindent\textbf{Engagement discriminators.}
We identify words that frequently appear in the titles of posts with different engagement levels.
Using a method based on the chi-squared test and contingency tables, we identify words that differ significantly in usage, hence discriminating between engaging and non-engaging posts. We lemmatize the titles, remove stop words, and split the posts into two groups (engaging vs. non-engaging).
Next, we identify the $100$ most commonly used words in each group and calculate chi-square ($\chi^2$) values from contingency tables to assess the statistical significance. 
Using this method, we identify $91$ words used with significant variation in engagement and non-engagement posts.
To ensure the relevance of these discriminative words, we sort identified discriminators by their occurrence frequency and select words that occur in at least $1$\% of posts, which results in the four most frequent engagement discriminators for engaging vs. non-engaging posts.
Finally, we create two new binary features indicating whether a post title contains a discriminator ($1$ for presence, $0$ for absence) specific to engagement or non-engagement posts.
In Figure \ref{fig:worclouds_both}, we depict the engagement discriminators as the word clouds categorized by engagement levels.
Mainly, we show that engaging posts have titles with ``cheese'', ``pizza'', or ``fry'', while non-engaging posts feature words such as ``rice'' or ``sauce.''
The prominence of words like ``cheese'' and ``pizza'' in posts with high engagement levels suggests that indulgent or popular foods may attract more attention.
Furthermore, words related to preparation methods, such as ``fry'' and ``smoked'' seem to play a critical role in capturing user interest.

\noindent\textbf{Visual features.} 
To extract visual features we compute the image embeddings for post images using CLIP (\textit{clip-ViT-L-14}) model \cite{radford_learning_2021}. The CLIP embeddings are $768$-dimensonal vectors that we reduce to five dimensions with UMAP \cite{mcinnes_umap_2020} for comparability with other feature groups and model interpretability.

\noindent\textbf{Control features.} 
We define control features as factors that typically influence user engagement in social media but are unrelated to food. In particular, we compute the following features: (i) user experience indicator (top $5$\% of most active users vs. the remaining users), (ii) indicator for the first, second, third, or fourth quartile of the day, (iii) weekend or weekday indicator, and (iv) pre-, during, or post-peak of the COVID-19 pandemic indicator. In addition, we use the post tag as another control variable. 

With control features, we account for various factors that could affect engagement apart from nutritional content and other food-related features such as visual appearance or textual indicators. For example, 
user activity levels vary between weekends and weekdays, influencing post frequency and audience size (cf. Figure \ref{fig:temporal}).
The COVID-19 pandemic altered user behavior on social media, making timing relative to the pandemic an important factor \cite{gligoric_population-scale_2022}. 
Posts by experienced users typically receive more engagement due to user familiarity with popular post attributes or their reputation in the subreddit \cite{rokicki_how_2017}.
Moreover, the visual content of an image strongly influences its popularity and engagement it receives \cite{khosla_what_2014}.
Finally, dividing the day into quartiles (six hours each) ensures balanced analysis across different times of the day.

\subsection{Experimental setup}
\label{sec:prediction}

We conduct a classification experiment on posts with engagement vs. posts without engagement. Using our five 
feature sets, we repeat predictions for all combinations of feature sets, always including control features. This results in $16$ different combinations. We divide our dataset into train ($80\%$) and test ($20\%$).
As the evaluation metric, we use the ROC-AUC score.

Using our training dataset with control features only and 5-fold cross-validation, we first optimize the hyperparameters for the XGBoost classifier. To that end, we combine randomized and grid search over these parameter values: $10$, $50$, $100$, $500$, $1,000$ and $5,000$ estimators; maximum tree depth of $1$, $2$, $3$, $4$, $10$, $15$; and learning rate of $0.01$, $0.1$, $0.2$, $0.3$, or $0.4$.
Randomized search allows us to estimate potential parameters quickly by randomly sampling combinations, facilitating a faster initial exploration. After identifying promising parameters, we define a range around these values for each hyperparameter and optimize them through grid search.
Ultimately, our final model for engagement prediction is configured with $70$ estimators, maximum tree depth of $2$, and learning rate of $0.22$.
Using the optimized parameters, we train the final XGBoost model on our training dataset for all different feature set combinations and evaluate the model on the test dataset with the ROC-AUC score. To estimate uncertainty in the test performance, we create $1,000$ bootstrap samples from the test dataset. 
Using those bootstrap samples, we calculate $95\%$ confidence intervals for the ROC-AUC score.

Finally, we estimate feature importance using SHAP values. SHAP (SHapley Additive exPlanations) \cite{lundberg_local_2020} values explain individual predictions of machine learning models by revealing contribution of each feature to the final prediction. By aggregating local explanations of each prediction, SHAP values offer an understanding of the global structure of the model, helping us understand overall impact of features on the predictions.

%\section{Results and Discussion}
\subsection{Results}

\begin{table}[t]
\centering
\caption{\textit{Results.} ROC-AUC engagement prediction scores for the control model (C) and models using Nutrition (N), Vision (V), Food Descriptors and Categories (F), and Engagement Discriminators (E). We report mean ROC-AUC with 95\% bootstrap confidence intervals and improvement over the control model $\Delta C$ (absolute and relative as percentage).}
\Description{Table showing ROC-AUC scores for every model.}
\resizebox{\columnwidth}{!}{
\begin{tabular}{@{}l c c c@{}}
\toprule
Model & ROC-AUC & 95\% CI & $\Delta C$ (\%) \\
\midrule
\multicolumn{4}{@{}l}{\textbf{Single feature set models}} \\
C (Control)                    & 0.644 & [0.612, 0.676] & ---  \\
C + Nutrition (N)              & 0.675 & [0.644, 0.703] & +0.031 (4.81\%) \\
C + Vision (V)                 & 0.753 & [0.724, 0.781] & +0.109 (16.89\%) \\
C + Food Descriptors and Categories (F)       & 0.647 & [0.616, 0.676] & +0.003 (0.47\%) \\
C + Engagement Discriminators (E) & 0.675 & [0.644, 0.705] & +0.031 (4.81\%) \\
\addlinespace
\multicolumn{4}{@{}l}{\textbf{Two feature sets models}} \\
C + N + V                      & 0.754 & [0.726, 0.779] & +0.110 (17.08\%) \\
C + N + F                      & 0.683 & [0.652, 0.713] & +0.039 (6.06\%) \\
C + N + E                      & 0.694 & [0.663, 0.722] & +0.050 (7.78\%) \\
\textbf{C + V + F}             & 0.755 & [0.726, 0.781] & +0.111 (17.23\%) \\
C + V + E                      & 0.753 & [0.726, 0.779] & +0.109 (16.89\%) \\
C + F + E                      & 0.680 & [0.651, 0.710] & +0.036 (5.59\%) \\
\addlinespace
\multicolumn{4}{@{}l}{\textbf{Three+ feature sets models}} \\
C + N + V + F                  & 0.748 & [0.720, 0.774] & +0.104 (16.15\%) \\
C + N + V + E                  & 0.751 & [0.724, 0.780] & +0.107 (16.61\%) \\
C + N + F + E                  & 0.695 & [0.664, 0.722] & +0.051 (7.92\%) \\
C + V + F + E                  & 0.750 & [0.723, 0.777] & +0.106 (16.46\%) \\
C + N + V + F + E              & 0.754 & [0.727, 0.782] & +0.110 (17.08\%) \\
\bottomrule
\end{tabular}
}
\label{table:auc_roc_scores}
\end{table}

We present our results in Table \ref{table:auc_roc_scores}, where we summarize our main findings, with ROC-AUC scores and their corresponding bootstrap confidence intervals for each model.

\noindent\textbf{Predicting engagement.}
Using only the control feature set, our classification model achieves a ROC-AUC score of $0.644$.
Adding nutritional attributes to the controls improves the score to $0.675$, or by $4.81$\%.
Including visual features results in the highest single feature set improvement, by $16.89$\%, raising the ROC-AUC to $0.753$.
When adding food descriptors \& categories, we observe only a small improvement ($0.47$\%) in performance (ROC-AUC of $0.647$), and when adding engagement discriminators, performance improves by $4.81$\% to $0.675$. 
Within the two feature sets models, the largest improvement ($17.23$\%) is achieved by combining vision and food descriptors \& categories, reaching a ROC-AUC score of $0.755$. Also, this is the best performing model overall. However, this result is not significantly different then the result obtained when combining nutrition and vision results (the second largest improvement overall), which improves by $17.08$\% over the baseline model and reaches a ROC-AUC of $0.754$.
Among models with three or more feature sets, the combination of all feature sets performs best, achieving the same score and improvement as the nutrition and vision model (ROC-AUC of $0.754$, $17.08$\% improvement).

\begin{figure*}[t]
    \centering
    \scalebox{1}{
    \begin{subfigure}[t]{0.49\linewidth} 
        \centering
        \includegraphics[width=\linewidth]{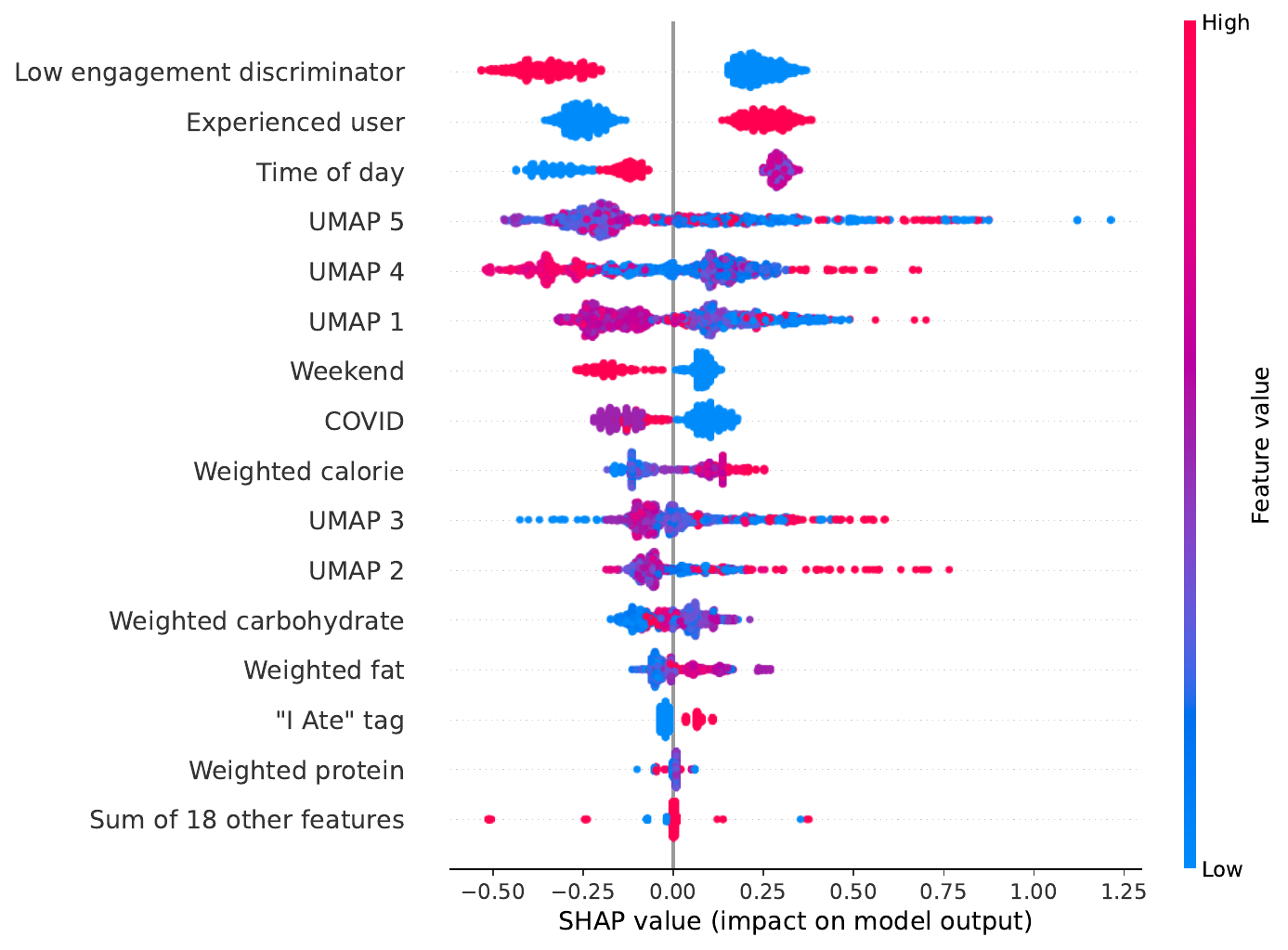}
        \caption{Feature impact on predicting high engagement.}
        \Description{Beeswarm chart of SHAP values for each feature.}
        \label{fig:shap_rq2}
    \end{subfigure}
    \hfill
    \begin{subfigure}[t]{0.43\linewidth}
        \centering
        \includegraphics[width=\linewidth]{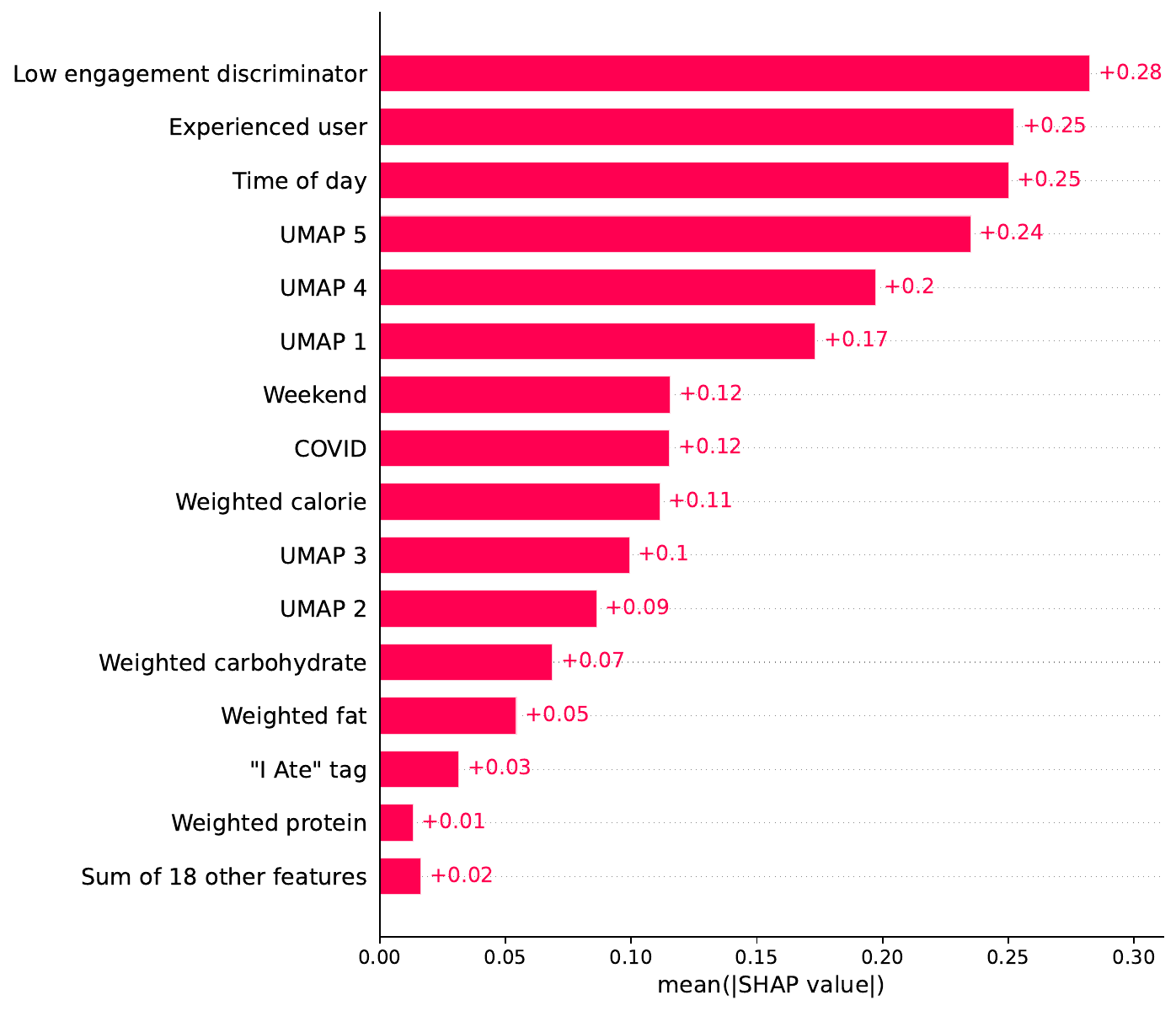}
        \caption{Overall feature importance.}
        \Description{Bar chart of mean SHAP values.}
        \label{fig:shap_bar_rq2}
        \end{subfigure}
    }
    \caption{
        \textit{SHAP visualizations for classifier predicting post engagement.} 
        SHAP values of features provide explanations for classifier predictions and allow us to understand which features contribute to prediction of high engagement. In (a), we present how the values of features impact the prediction, while in (b) we present the overall feature importance.
        While some of the control (user experience and time of the day), textual (words appearing frequently in titles of non-engaging posts), and visual (UMAP 1 to UMAP 5) features have a substantial impact on the classifier predictions, nutritional features are among the most important ones, with the calorie density being the 9th most important feature overall (out of 33 features). Specifically, higher calorie density increases the probability for prediction of engagement. Similarly, a higher fat value has a positive influence on the engagement prediction, while protein and carbohydrate effects are more nuanced.
        }
    \label{fig:shap_all_rq2}
    \Description{Two charts provide a SHAP analysis of a classifier predicting high engagement for r/Food posts across 33 different features. Beeswarm chart shows feature impact on prediction, while bar chart shows overall feature importance.}
\end{figure*}

\noindent\textbf{Feature importance with SHAP values.}
To better understand the associations between individual features and user engagement, we calculate SHapley Additive exPlanations (SHAP) values for the model that includes all feature sets. In particular, SHAP values capture the contribution of individual features to the prediction of high engagement. Specifically, a positive SHAP value is associated with an increase in the probability of a positive prediction (engagement), while a negative SHAP value is related to a decrease in that probability.
For example, in Figure \ref{fig:shap_rq2} we illustrate how individual features influence predictions of whether a post will receive engagement by plotting the SHAP values across posts.
In particular, we observe how different feature values affect the model's prediction.
Hence, each point represents a SHAP value for an individual post given its corresponding feature value.
The SHAP values are given on the x-axis, reflecting their impact on the model's output.
Specifically, positive SHAP values push the prediction towards engagement, while negative values reduce the probability of the positive class.
The color gradient represents the feature value, with blue indicating lower feature values and red indicating higher values. For example, blue points in the ``Weighted calorie'' row of Fig. \ref{fig:shap_rq2} represent lower calorie while red points represent higher calorie density.
Additionally, Figure \ref{fig:shap_bar_rq2} displays the mean of absolute SHAP values across the range of possible feature values. 
While the plot does not differentiate the direction of the feature's SHAP values, it depicts the overall feature importance in the prediction model.

\noindent\textbf{Role of nutritional content in engagement prediction.}
In Figure \ref{fig:shap_rq2} and \ref{fig:shap_bar_rq2}, we present associations between individual features and post engagement. We observe that higher calorie density is positively associated with engagement (red points on the positive SHAP axis for weighted calorie), with a mean absolute SHAP value of $0.11$ indicating that, on average, the calorie density changes the prediction probability for high engagement by 11\%.
Although the absence of low resonance discriminators, high user experience, and temporal features (the latter two being non-food related control features), and visual features (denoted as UMAP 1 to UMAP 5 in the diagrams),  have very strong influence on engagement prediction, the calorie density is clearly an important feature ranked as the 9th (out of 33) most important feature overall.
Hence, posts featuring high-calorie meals are more likely to receive higher levels of engagement from the community even after accounting for the influence of the control features and other feature sets. On the other hand, low-calorie meals tend to be associated with non-engaging posts (red points on the negative SHAP axis for weighted calorie).
Similarly, SHAP values suggest that high fat content is also linked with positive engagement prediction.
In contrast, both carbohydrate and protein content have a more nuanced correlation with engagement.
While low carbohydrate density is associated with non-engaging posts, engaging posts contain both high- and mid carbohydrate meals.
Conversely, high protein density is associated with non-engaging posts, low protein content tends to appear more likely in posts that receive engagement.

\subsection{Discussion}

Our findings on the relationship between nutritional content and engagement in food-related social media posts provide key insights into user behavior, as well as the context and content of engaging posts. First, including the nutritional content as a feature set in our engagement prediction models significantly enhances the baseline model's classification accuracy, suggesting a strong predictive power of these nutritional features for engagement. Additionally, we uncover the direction of this strong association: more calorie-dense meals increase the prediction probability for user engagement. This influence of calorie content aligns with prior research suggesting that users are more drawn to calorie-dense meals \cite{pancer_content_2022}.
Posts with higher calorie content consistently demonstrate higher SHAP values, emphasizing their role in engagement prediction.
The interaction between calorie and fat density further corroborates this, as posts featuring both high-calorie and high-fat meals are typically more likely to reach high engagement.
Conversely, while low carbohydrate content is linked to posts with low engagement, resonant posts feature meals with various carbohydrate levels. 
Moreover, we hypothesize that the nuanced effect of protein suggests high-protein meals appeal to a niche audience and, hence, not resonate well enough with a broader user community.

However, several other features including visual and some of the textual features are also strongly associated with the user engagement with posts. For example, in our dataset, the visual features exhibit the strongest predictive performance for user engagement, which is a common occurrence on social media \cite{khosla_what_2014, hessel_cats_2017}. 
Specifically, visual features, food descriptors \& categories, as well as title words discriminating between engaging and non-engaging posts all directly capture the type of the food that users present, and hence, already encapsulate some of information about the nutritional content of that food due to strong correlations between these features and nutritional content. For example, meals such as pizza or fast food are more calorie-dense than salads or other healthy dishes. However, nutritional content can have different sources not restricted only to the meal names as the calorie density of the meal may be the result of the way how the meal is prepared (e.g., fried vs. cooked), the particular composition of the ingredients, or simply a large portion size visible in the posted image. Hence, while nutritional content is correlated with other features, it also combines various information sources in unique ways, adding important predictive information to our models. 
In other words, high-calorie meals are typically associated with higher engagement levels regardless of the source of their calorie density. 

This association between nutritional content and the user engagement remains stable even after controlling for multiple features such as user experience, time of the posting, food descriptors and categories, or significant words used in the titles.
For example, user experience appears as a critical feature strongly related to engagement \cite{bakshy_everyones_2011}.  
Posts by more experienced users have a higher likelihood of engagement, confirming findings from studies on social media websites such as Twitter, where contributions by long-term users were more likely to receive responses \cite{suh_want_2010}.
This phenomenon may be related to the community perceiving content from experienced users as being of higher quality, or to the experienced users being able to understand the community and their expectations better than inexperienced ones. Further, the timing of the posts is also significantly related to engagement. 
While overall post and comment volume increased after the onset of COVID-19, potentially due to increased digital screen time during lockdowns \cite{wong_digital_2021}, the prediction probability for high engagement posts increases for posts before COVID-19, suggesting a more uniform engagement distribution post COVID-19.
Moreover, posts made later in the day or during weekends were less likely to engage users, agreeing with the findings that weekday posts during busier hours attract more interaction \cite{wahid_social_2020, hanifawati_managing_2019}.
Finally, our findings align with studies indicating that captions and post titles significantly influence engagement \cite{hessel_cats_2017, chen2021drives}, reinforcing the importance of carefully crafting titles and captions to resonate with audiences.

\noindent\textbf{Additional sensitivity analysis.} Apart from the robustness checks that we described earlier, which include the use of several similarity thresholds for nutritional content calculation, and different comment counts (up to five) for identification of low-engaging posts, we conduct an additional classification experiment to further assess the robustness of our results.
Specifically, we use a total of $207,248$ posts including posts with at least one comment ($103,624$) and an equal number of randomly sampled posts with no comments, and train another XGBoost model to predict whether a post will receive comments or not.
In these experiments, the model with control features achieves an ROC-AUC score of $0.584$. Adding nutritional content improves the score to $0.597$ (improvement of $2.23$\%), and using all features increases it to $0.617$ ($5.65$\% improvement). This model with all features is the best performing model, alongside with the model that uses nutrition content, visual features and engagement discriminators.
These results are comparable to the initial experiment, with a lower overall performance due to a weaker separation between the classes. Additionally, the relative ranking in SHAP values of features slightly shifts.
However, calorie density still ranks as the $9$th important feature (out of 33), with a mean absolute SHAP value of $0.05$.
On the other hand, both protein and fat have polarizing effects. Low protein, or high fat values, are either boosting or decreasing the probability of predicting engagement.

\noindent\textbf{Limitations.}
Even though Reddit supports more authentic behavior due to its anonymity, it comes with several limitations.
First, we miss the detailed user demographics, individual interests, or nutritional goals. 
Second, we do not account for bots, which users can easily create \cite{long2017could}, and which can influence user engagement.
Third, Reddit's algorithm that curates feeds may influence user engagement with specific, assumed-relevant posts.

Furthermore, even though we include UMAP reduced visual features, we do not investigate their interpretation.
Additionally, we measure engagement by the number of comments regardless of sentiment, and future work could explore the relationship between nutritional content and qualitative engagement patterns.

Although we use pre-trained embeddings, a robust similarity threshold, and similarity-weighted aggregation to estimate nutritional content, we acknowledge potential inaccuracies. We estimate nutritional densities rather than total amounts, and ingredient ratios may vary.
Additionally, common meals, such as pizza, can have numerous variations, and users may not feel the need to specify these differences in the title, as they accompany their title with a picture.
Since our method relies solely on the title, our approach may overlook information valuable for the accuracy of the estimation.
Incorporating visual features into the nutrition estimation process could provide deeper insights and help mitigate the lack of detail about specific ingredients or portion sizes in common meals. We leave this estimation as a potential avenue for future work.

Moreover, our work is a large-scale study of a single, although large, community (i.e., Reddit's r/Food). 
While we believe that the amounts of data (almost 600,000 posts) and user base (24 million) are sufficient, we acknowledge potential sample bias in users of this community.
Therefore, our findings might not necessarily generalize to other communities. 
However, we see this as an opportunity to extend our work to other social media platforms that garner a large number of users, such as Instagram.

Finally, we caution that our work indicates an associative link between nutritional content and different levels of engagement, and does not establish causality.
Albeit we control for several confounding features, which makes the evidence we find for this link stronger, our observational setup lacks the structure needed for causal inference. 

\section{Conclusion}
\noindent\textbf{Summary.}
In this work, we explored the association between the nutritional content of food-related posts on Reddit's r/Food community and user engagement.
By estimating the nutritional content with an innovative embedding-based method just from post titles and analyzing almost $600,000$ posts, we uncover that nutritional information is predictive of engaging posts resonating well with the community.
Our findings suggest that posts featuring calorie-dense meals are positively associated with higher user engagement, even after controlling for non-food-related factors.
This work expands on previous studies by focusing on only textual information for estimating nutritional content and by conducting a large-scale analysis of the relationship between calorie and macronutrient density and engagement, highlighting the role of nutritional content.

\noindent\textbf{Implications.}
The underlying study has several implications.
First, it provides insights into the driving factors behind user engagement with food-related content online. We uncover patterns that drive user behavior by employing large-scale analysis and exploring the intersection of technology, nutrition, and social engagement. More specifically, we provide information on the nutritional and general characteristics of posts that users engage with.
Next, the explainability of our models allows us to structure the posts that are more likely to garner engagement. This allows the design of impactful online initiatives aimed at promoting healthy eating choices.
Furthermore, the improved estimation of nutritional content solely from textual description provides an accessible and scalable tool for dietary education, offering individuals a way to understand the profile of their meals.
These applications might encourage users to make informed dietary and health decisions.
Finally, our findings contribute to the broader discourse on how technology shapes social outcomes and practices, especially in the area of health.

\noindent\textbf{Future Work.}
In future work we can further improve the calorie estimation method by accounting for the visual features. Also, extending our analysis to other social media platforms can provide a comparison of engagement patterns across diverse communities.
Finally, while our study analyzes the correlation between nutritional content and engagement, future studies could also explore causal relationships, potentially using experimental and quasi-experimental designs.

%\begin{acks}
%To Robert, for the bagels and explaining CMYK and color spaces.
%\end{acks}

\bibliographystyle{ACM-Reference-Format}
%\bibliography{sample-base}
\bibliography{references-new}

%\appendix


\end{document}